\documentclass[aps,prd,floatfixy,superscriptaddress,showpacs,showkeys]{revtex4}[14pt]
\usepackage{amssymb}
\usepackage{amsmath}
\usepackage{amsfonts}
\usepackage{epsfig}
\usepackage{graphicx}
\usepackage{tabularx}
\usepackage{latexsym}
\usepackage{subfigure}
\usepackage{verbatim}
\usepackage{color}
\usepackage{braket}
\usepackage{multirow}
\usepackage{dcolumn}
\usepackage{bm}
\newcommand{\seq}{\begin{subequations}}
\newcommand{\sen}{\end{subequations}}
\newcommand{\eq}{\begin{eqnarray}}
\newcommand{\en}{\end{eqnarray}}

\def\ds{D^{\ast  0}}

\def\L2{\Lambda^2}

\def\eq{\begin{eqnarray}}
\def\en{\end{eqnarray}}

\usepackage[normalem]{ulem}

\renewcommand\sout{\bgroup \color{red} \ULdepth=-.5ex \ULset}

\def\ds{D^{\ast  0}}

\def\L2{\Lambda^2}

\def\eq{\begin{eqnarray}}
\def\en{\end{eqnarray}}

\def\L{{\cal L}}

\def\ds{d^*}

\begin{document}

\title{\bf \boldmath On the charge distribution of $d^*(2380)$}
\vspace{1.5cm}
\author{Yubing Dong}
\affiliation{Institute of High Energy Physics, Chinese Academy of
Sciences, Beijing 100049, China}
\affiliation{Theoretical Physics
Center for Science Facilities (TPCSF), CAS, Beijing 100049, China}
\affiliation{School of Physical Sciences, University of Chinese
Academy of Sciences, Beijing 101408, China}

\author{Fei Huang}
\affiliation{School of Physical Sciences, University of Chinese
Academy of Sciences, Beijing 101408, China}

\author{Pengnian Shen}
\affiliation{College of Physics and Technology, Guangxi Normal
University, Guilin  541004, China}
\affiliation{Institute of High
Energy Physics, Chinese Academy of Sciences, Beijing 100049, China}
\affiliation{Theoretical Physics Center for Science Facilities
(TPCSF), CAS, Beijing 100049, China}

\author{Zongye Zhang}
\affiliation{Institute of High Energy Physics, Chinese Academy of
Sciences, Beijing 100049, China} \affiliation{Theoretical Physics
Center for Science Facilities (TPCSF), CAS, Beijing 100049, China}
\affiliation{School of Physical Sciences, University of Chinese
Academy of Sciences, Beijing 101408, China}

\date{\today}

\begin{abstract}
We calculate the charge distributions of $d^*(2380)$. Two different
interpretations of the $d^*$ are considered for a comparison. One is a
compact explanation with coupled $\Delta\Delta+CC$ two-channel approximation
in the chiral constituent quark model. Another is a resonance state of $D_{12}\pi$.
The remarkable differences of the charge distributions in the two pictures are shown
and it is expected that the future experiments may provide a clear test for
the different theoretical interpretations.
\end{abstract}

\pacs{13.25.Gv, 13.30.Eg, 14.40.Rt, 36.10.Gv}

\keywords{$\ds(2380)$, Chiral quark model, Charge distributions, $d^*(2380)$, $D_{12}\pi$}

\maketitle

\section{Introduction}

$d^*(2380)$ is a new resonance recently observed by CELSIUS/WASA and
WASA@COSY Collaborations~\cite{CELSIUS-WASA,CELSIUS-WASA1}. It was
found in the analysis of double pionic fusion channels $pn\to
d\pi^0\pi^0$ and $pn\to d\pi^+\pi^-$ when the ABC effect~\cite{ABC}
and the analyzing power $A_y$ of the neutron-proton scattering data
were studied. It is argued that the observed structure cannot be
simply understood by either the intermediate Roper excitation
contribution or by the t-channel $\Delta\Delta$ process,
Refs.~\cite{CELSIUS-WASA,CELSIUS-WASA1} proposed an assumption of
existing a $d^*$ resonance whose quantum number, mass, and width are
$I(J^P)=0(3^+)$, $M \approx 2370$ MeV and $\Gamma \approx 70$ MeV
(see also their recent paper~\cite{Bashkanov}, the averaged mass and
width are $M \approx 2375$ MeV and $\Gamma \approx 75$ MeV,
respectively). Since the  baryon number of $d^*$ is
2, it would be treated as a dibaryon, and could be explained by
either "an exotic compact particle" or "a hadronic molecule
state"~\cite{CERN}. Moreover, since the observed mass of the $d^*$
is about 80~MeV below the $\Delta\Delta$ threshold and about
$70~MeV$ above the $\Delta\pi N$ threshold, the threshold (or cusp)
effect may not be so significant as that in the XYZ particles (see
the review of XYZ particles ~\cite{Chen:2016qju} for example). Thus,
understanding the internal structure of $d^*$ would be of great interest.\\

The existence of such a non-trivial six-quark configuration with
$I(J^P)=0(3^+)$ (called $d^*$ lately) has triggered a great
attention and has intensively been studied in the literature even
before the COSY's
discovery~\cite{Dyson,Thomas,Oka,Wang,Yuan,Kukulin}. In fact, after
the experimental observation of the $\ds$, there are mainly three
types of model explanation for its nature in the market. The first
one~\cite{Wang1} proposed a $\Delta\Delta$ resonance structure and
obtained a binding energy of about 71 MeV (namely $M_{d^*}=2393$
MeV) and a width of about $150$ MeV.  The second one~\cite{Gal}
announced a broad resonant structure of $D_{12}\pi$ for $\ds$, whose
resonant pole is around ($2363\pm 20$)~+{\it i} ($65\pm 17$) MeV.
The third one~\cite{Huang}, following our previous
prediction~\cite{Yuan}, suggested a dominant hexaquark structure for
$\ds$, with a mass of about $2380-2414$ MeV and a width of about
$71$ MeV, respectively~\cite{Huang,Dong,Dong1,Dong2017}. From the
results of those models, one has three observations: (1) All
proposed models can reproduce a right mass of $\ds$; (2) Only can
last two models provide a total width for $\ds$ which compatible
with the observed data; (3) only can the last model give all partial
decay widths whose values agree with the observed data quite well.
Even more, in terms of the third model (hexaquark dominant model),
the predicted decay width of the single pion decay mode of $d^*\to
NN\pi$ is about "1~MeV" which is much smaller than the double-pion
decay widths~\cite{Dong2017}. Here, we would particularly mention
that this value is much smaller than that with the second scenario
of $D_{12}\pi$, but agrees with the experimental observation very
recently~\cite{Clement2017}. All the  outcomes from the third
scenario support the idea that $\ds$ is probably a compact six-quark
dominated exotic state due to its large $CC$ component by Bashkanov,
Brodsky and Clement \cite{Brodsky}. Therefore, the model decay width
of the $\ds \to NN\pi$ process could be one of the way to justify
the structure of $\ds$, namely the width would be sensitive to the
structure of $\ds$. Although we have this weapon in hand, we are now
still facing the problem that there is any additional way to justify
different structure models.
The general review on the dibaryon studies can be found in Ref.~\cite{Clement}.\\

It is known that the electromagnetic probe is one of the most useful
tools to test the internal structure of a complicated system. For
example, the electromagnetic form factors of the nucleon show the
charge and magnetron distributions of a nucleon. The slops of the
charge and magnetic distributions at original gives the charge and
magnetic radii of the system. The precise measurement of the proton
charge radius provides a criteria for different model calculations.
For the spin-1 particle, like a deuteron or a $\rho$- meson, the
charge, magnetic and quadrupole form factors tell the intrinsic
structures as well, like charge and magnetron distributions and the
quadrupole deformation of the system. Therefore, the
form factors of $\ds$, for instance the charge distribution, might
also be a physical quantity for discriminating the structure of
$\ds$. In this work, the charge distribution of the new resonance
$d^*(2380)$ will be discussed with two different theoretical
scenarios  for comparison. One is a
system with a single $\Delta\Delta$ structure or a
coupled $\Delta\Delta+CC$ structure in our chiral
constituent quark model. The other one is a resonant
system of $D_{12}\pi$. Moreover, the $d^*(2380)$ is a spin-3
particle, it has $2S+1=7$ form factors. A detailed discussion of all
the seven form factors is beyond the scope of this work and will be
given elsewhere. Here we only concentrate on its charge distribution
and the charge
radius of the $d^*(2380)$ in the two scenarios.\\

This paper is organized as follows. In Sect. II, a brief
discussion about the electromagnetic form factors
of particles with spin-1/2, spin-1, and spin-3 will be shown.
An explicit calculation of the charge
distribution of $d^*$ with two scenarios is given in
Sect. III. Sect. IV is devoted to a short summary.\\

\section{Electromagnetic form factors}

The study of the electromagnetic form factors of the nucleon
(spin-1/2) is of great interest because it can tell {\color{blue}us}
the information about the charge and magnetron distributions of a
nucleon. In the one-photon approximation, the electromagnetic
current of a nucleon is \eq <N(p')\mid J^{\mu}_N\mid
N(p)>=\bar{U}_N(p')\Big
[F_1(Q^2)\gamma^{\mu}+i\frac{\sigma^{\mu\nu}q_{\nu}}{2M_N}F_2(Q^2)\Big
]U(p), \en where $M_N$ is nucleon mass, $q=p'-p$ is momentum
transfer, $Q^2=-q^2$, and $F_1(Q^2)$ and $F_2(Q^2)$ are the Dirac
and Pauli form factors, respectively. These two form factors relate
to the electric and magnetic form factors \eq G_E(Q^2)=F_1(Q^2)-\eta
F_2(Q^2), ~~~~~G_M(Q^2)=F_1(Q^2)+F_2(Q^2), \en with
$\eta=Q^2/4M_N^2$. The normalization conditions of the two form
factors for the proton and neutron are
$G_E^p(0)=1$, $G_E^n(0)=0$, $G_M^p=2.79$, and $G_M^n=-1.91$, respectively.\\

In the Breit frame, we have $q^{\mu}=(0,\vec{q})$, $p'^2=p^2=M_N^2$,
$\vec{p}=-\vec{p}^{~'}=-\frac12\vec{q}$, and
$p_0=p_0'=E=M_N(1+\eta)^{1/2}$. Then, we obtain~\cite{Barik} \eq
<N(\vec{q}/2)\mid J^{0}_N\mid
N(-\vec{q}/2)>&=&(1+\eta)^{-1/2}\chi_{s'}^+\chi_sG_E(Q^2)\\
\nonumber <N(\vec{q}/2)\mid \vec{J}_N\mid
N(-\vec{q}/2)>&=&(1+\eta)^{-1/2}\chi_{s'}^+\frac{\vec{\sigma}
\times\vec{q}}{2M_N}\chi_sG_M(Q^2).
\en
Clearly, the electric form factor $G_E(Q^2)$ is directly related
to the matrix element of $J_N^0$.\\

For a spin-1 particle, like deuteron, it contains three form
factors. In  the one-photon approximation, the electromagnetic
current is \eq J_{jk}^\mu(p',p)={\epsilon}^{'*\alpha}_j(p')
S^\mu_{\alpha\beta} {\epsilon}^{\beta}_k(p) \en
where $\epsilon$ and $\epsilon'$ stand for the polarization vectors
of the incoming and outgoing deuterons, $i$ and $k$ are the polarizations of
the two deuterons, and
\begin{eqnarray}
\label{eq:currD}
S_{\alpha \beta}^{\mu}=-\Big [G_1(Q^2)g_{\alpha \beta}-G_3(Q^2)
\frac{Q_{\alpha}Q_{\beta}}{2 m_D^2}\Big ] P^\mu
-G_2(Q^2)(Q_\alpha g_\beta^\mu-
Q_\beta g_\alpha^\mu) \ ,
\end{eqnarray}
where $P=p'+p$. The three form factors $G_{1,2,3}(Q^2)$ relate to
the charge $G_C(Q^2)$, magnetic $G_M(Q^2)$ and quadrupole form
factors $G_Q(Q^2)$ as~\cite{Barik}
\begin{eqnarray}
G_C(Q^2)&=&G_1(Q^2) + \frac{2}{3}{\eta}_D G_2(Q^2) \ ,
~~~~~G_M(Q^2)=G_2(Q^2) \ , \ \\
G_Q(Q^2)&=&G_1(Q^2) - G_2(Q^2) + (1+\eta_D)G_3(Q^2)\ , \
\end{eqnarray}
with $\eta_D=Q^2/4m_D^2$ and $M_D$ is the deuteron mass. The charge,
magnetic and quadrupole form factors are normalized to $G_C(0)=1$,
$G_M(0)=\frac{m_D}{m_N}\mu_d=1.714$, and $G_Q(Q^2)=m_D^2 Q_d=25.83$,
respectively. In the Breit frame, one may also see that the charge
form factor of the deuteron $G_C(Q^2)$ can be obtained by direct
calculating the matrix element $\frac13 \sum_{\lambda}<p',
\lambda \mid J^0\mid p, \lambda>$.\\

For the $d^*(2380)$ particle, since its spin is 3,
it has 2s+1=7 form factors. Its field can be expressed as
$\epsilon_{\alpha\beta\gamma}$ a rank-3 tensor which is traceless.
Clearly, $\epsilon_{\alpha\alpha\beta}=0$,
$\epsilon_{\alpha\beta\gamma}=\epsilon_{\beta\alpha\gamma}$, and
$p^{\alpha}\epsilon_{\alpha\beta\gamma}=0$. In the one-photon
exchange approximation, the general form of the electromagnetic
current of a $3^+$ particle is
\eq
{\cal J}^{\mu}=\big
(\epsilon^*\big )^{\alpha'\beta'\gamma'}(p'){\cal
M}^{\mu}_{\alpha'\beta'\gamma',\alpha\beta\gamma}\epsilon^{\alpha\beta\gamma}(p)
\en
and the matrix element \eq {\cal
M}^{\mu}_{\alpha'\beta'\gamma',\alpha\beta\gamma}&=& \Bigg
[G_1(Q^2){\cal P}^{\mu}\Big [g_{\alpha'\alpha}\Big
(g_{\beta'\beta}g_{\gamma'\gamma}+g_{\beta'\gamma}g_{\gamma'\beta}\Big
)+permutations\Big ]\\ \nonumber &&+G_2(Q^2){\cal P}^{\mu}\Big
[q_{\alpha'}q_{\alpha}\big [g_{\beta'\beta}g_{\gamma'\gamma}+
g_{\beta'\gamma}g_{\gamma'\beta}\big ]+permutations\Big ]/(2M^2)\\
\nonumber &&+G_3(Q^2){\cal P}^{\mu}\Big
[q_{\alpha'}q_{\alpha}q_{\beta'}q_{\beta}g_{\gamma'\gamma}+permutations\Big
]/(4M^4)
\\ \nonumber
&&+G_4(Q^2){\cal
P}^{\mu}q_{\alpha'}q_{\alpha}q_{\beta'}q_{\beta}q_{\gamma'}q_{\gamma}/(8M^6)\\
\nonumber &&+G_5(Q^2)\Big [ \Big
(g_{\alpha'}^{\mu}q_{\alpha}-g_{\alpha}^{\mu}q_{\alpha'}\Big)
\Big(g_{\beta'\beta}g_{\gamma'\gamma}+g_{\beta'\gamma}g_{\beta'\gamma}\Big
)+permutations \Big ]\\ \nonumber &&+G_6(Q^2)\Big [ \Big
(g_{\alpha'}^{\mu}q_{\alpha}-g_{\alpha}^{\mu}q_{\alpha'}\Big )
\Big(q_{\beta'}q_{\beta}g_{\gamma'\gamma}+q_{\gamma'}q_{\gamma}g_{\beta'\beta}
+q_{\beta'}q_{\gamma}g_{\gamma'\beta}+q_{\gamma'}q_{\beta}g_{\gamma\beta'}\Big
)\\ \nonumber &&~~~~~~~~~~~~~~+permutations\Big ]/(2M^2)\\ \nonumber
&&+G_7(Q^2)\Big [\Big
(g_{\alpha'}^{\mu}q_{\alpha}-g_{\alpha}^{\mu}q_{\alpha'}\Big
)q_{\beta'}q_{\beta}q_{\gamma'}q_{\gamma}+permutations\Big ]/(4M^4)
\Bigg ]
\en
where $M$ is the mass of $d^*$, ${\cal P}=p'+p$, and $G_{1,2,3,4,5,6,7}(Q^2)$ are the
seven form factors, respectively. The gauge invariant condition
\eq
q_{\mu}{\cal M}^{\mu}_{\alpha'\beta'\gamma',\alpha\beta\gamma}=0
\en
is fulfilled as well as the time-reversal invariance. The
combinations of the above seven form factors $G_{1,2,...7}(Q^2)$ can
give the physical form factors of $d^*$ such as
the charge, magnetic, quadrupole as well as other
higher-order multipole form factors. The normalization of all the
form factors are unknown except for the charge form factor of
$G_c(0)=1$. It should be mentioned that the discussion of all the
seven form factors is beyond the scope of this paper, and it will
appear elsewhere. \\

In analogy to the spin-1/2 nucleon and spin-1 deuteron cases, we
assume that the charge distribution of the spin-3 particle,
$d^*(2380)$, is also directly relate to the matrix element of ${
J}^0$ in the Breit frame. For the six-quark system, we just consider
the quark-quark-photon current
\eq
J^0=\sum_{i=1}^6e_i\bar{q}_i\gamma^0q_i=\sum_{i=1}^6 j^0_i.
\en
Thus, we may calculate the matrix element of $J^0$ and determine the
charge distributions of the $d^*(2380)$ in the form of
\eq
G_E^{d^*}(Q^2)=\frac{1}{7}\sum_{m_{d^*}=-3}^{3}<p',m_{d^*}\mid
J^0 \mid p,m_{d^*}>.
\en

\section{Calculations of the $d^*$ charge distribution in two scenarios}

\subsection{Scenario A: Hexaquark dominant structure}

Here, we only concentrate on the charge distribution of $d^*(2380)$.
As mentioned in Refs.~\cite{Huang,Dong}, our model wave function for
$d^*(2380)$ is obtained by dynamically solving the bound-state RGM
equation of the six quark system in the framework of the extended
chiral $SU(3)$ quark model, and then successively projecting the
solution onto the inner cluster wave functions of the $\Delta\Delta$
and CC channels. The resultant wave function of $d^*$ can finally be
abbreviated to a form of \eq \label{wf1} |\Psi_{d^*(2380)}\rangle
&=&~\alpha|~\Delta\Delta~\rangle_{(SI)=(30)}~+~
\beta|~CC~\rangle_{(SI)=(30)}\\ \nonumber
&=&|~[~\phi_{\Delta}(\vec{\xi}_1,\vec{\xi}_2)~\phi_{\Delta}(\vec{\xi}_4,\vec{\xi}_5)~
\chi_{\Delta\Delta}(\vec{R})~\zeta_{\Delta\Delta}~ +~
\phi_{C}(\vec{\xi}_1,\vec{\xi}_2)~\phi_{C}(\vec{\xi}_4,\vec{\xi}_5)~
\chi_{CC}(\vec{R}))~\zeta_{CC}~]~\rangle_{(SI)=(30)},\\ \nonumber
\en where $\alpha$ and $\beta$ are the fractions of the
$\Delta\Delta$ and $CC$ components in $\ds (2380)$, $\phi_{\Delta}$
and $\phi_{C}$ denote the inner cluster wave functions of $\Delta$
and $C$ (color-octet particle) in the coordinate space,
$\chi_{\Delta\Delta}$ and $\chi_{CC}$ represent the channel wave
functions in the $\Delta\Delta$ and CC channels (in the single
$\Delta\Delta$ channel case, the $CC$ component is absent), and
$\zeta_{\Delta\Delta}$ and $\zeta_{CC}$ stand for the spin-isospin
wave functions in the hadronic degrees of freedom in the
corresponding channels, respectively~\cite{Huang}. It should be
specially mentioned that in such a $\ds$ wave function, two channel
wave functions are orthogonal to each other and contain all the
totally
anti-symmetrization effects implicitly~\cite{Huang}.\\

Unlike in calculations of  the decay processes of
$d^*\to d\pi\pi$, $d^*\to NN\pi\pi$, and $d^*\to NN\pi$ where the
$CC$ component does not contribute to the widths, here
in the calculation of the charge distribution of the $d^*$
both the $CC$  and $\Delta\Delta$ components contribute.
Considering that $\Delta$ and $C$ are antisymmetric then the charge
distribution is
\eq
G^{A}_E(Q^2)&=&<\ds({\cal P}')\mid \sum_{i=1}^6 j_i^0\mid
\ds({\cal P})>\\ \nonumber &=&3<\ds({\cal P}')\mid \big
(e_3j_3^0+e_6j_6^0\big )\mid \ds({\cal P})>,
\en
where the superscribe "A" stands for the scenario A,
$\vec{{\cal P}}'-\vec{{\cal P}}=\vec{q}$, $Q^2=\vec{q}^{~2}$,
and $e_{3,6}=\frac16+\frac{\tau^z_{3,6}}{2}$. Then
\eq
G^A_E(Q^2)=3\Bigg[\alpha^2
(I_3^{\Delta}+I_6^{\Delta}){\cal{O}}_{\Delta}
{\cal O}_{\chi_{\Delta}}
+\beta^2(I_3^{C}+I_6^{C}){\cal{O}}_{C}
{\cal O}_{\chi_{C}}\Bigg ].
\en
where
${\cal{O}}_{\Delta,C}$ denote the overlaps of the wave functions of
the 3-rd quark (or 6-th quark) which bombarded by photon in $\Delta$
and C, and ${\cal O}_{\chi_{\Delta}}$ and ${\cal O}_{\chi_{C}}$
represent the contributions from the $\Delta\Delta$ and $CC$ channel
wave functions, respectively. $I_{3,6}^{\Delta,C}$ can be calculated
by
\eq
I_{3,6}^{\Delta}~&=&~_{(I,I_z)}\langle ~\Delta\Delta~|
~e_{3,6}~|~\Delta\Delta~ \rangle_{(I,I_z)},\\ \nonumber
I_{3,6}^{C}~&=&~_{(I,I_z)}\langle ~CC~| ~e_{3,6}~|~CC~
\rangle_{(I,I_z)}.\\ \nonumber
\en
Finally, one obtains
\eq\label{Gddcc}
G^A_E(Q^2)=\Bigg[{\alpha}^2 \exp\Big
[-\frac{b_{\Delta}^2q^2}{6}\Big ] {\cal O}_{\chi_{\Delta}}
+{\beta}^2 \exp\Big[-\frac{b_{c}^2q^2}{6}\Big ] {\cal
O}_{\chi_{c}}\Bigg ].
\en
where $b_{\Delta,C}$ are the
size-parameters of the $\Delta$ and $C$ systems. As has been
discussed explicitly in Ref.~\cite{Dong}, the channel
wave function in the
$\Delta\Delta$ channel is written in terms of Gaussian-like wave
functions as
\eq \label{rwf1}
\chi(r)_{\Delta\Delta}=\sum_{m=1}^4\frac{c_m}{\sqrt{4\pi}}
\exp\Big(-\frac{r^2}{2b_m^2}\Big),
\en
where $c_m$ and $b_m$ can be
determined by fitting the channel wave function in this
form to our projected model wave function calculated before.
The normalization condition of the wave function is $\int d^3r \mid
\chi(r)\mid^2=1$. Thus
\eq
{\cal O}_{\chi_{\Delta}}(Q^2)=<\chi_{\Delta\Delta}(\vec{k}+\vec{q}/2)
\mid\chi_{\Delta\Delta}(\vec{k})>=\sum_{mn}c_mc_n\frac{b_m^3b_n^3}{4\pi}
\Big (\frac{2\pi}{b_m^2+b_n^2} \Big )^{3/2}exp\big
[-\frac{b_m^2b_n^2}{8(b_m^2+b^2_n)}\vec{q}^2\big ].
\en
For the $CC$ component, the
channel wave function is dominated by the single
S-wave Gaussian function
\eq
\chi_{CC}(r)=\Big (b_c^r\sqrt{\pi}\Big
)^{-3/2}\exp\Big [-r^2/2(b_c^r)^2\Big ].
\en
Similarly, the contribution from the $CC$ component can be calculated by
\eq
{\cal{O}}_{\chi_{C}}(Q^{2})=<\chi_{CC}(\vec{k}+\vec{q}/2)\mid\chi_{CC}(\vec{k})>=\exp
\Big[-\frac{(b_c^r)^2}{16} \vec{q}^2 \Big].
\en

\subsection{Scenario B: $D_{12}\pi$ structure}

We can also calculate the wave function of $D_{12}$ with (I=1, S=2)
by using our chiral SU(3) constituent quark model. The
obtained mass is about $M_N+M_{\Delta}-\epsilon\sim 2175-\epsilon ~MeV$ ($\epsilon$ is binding energy).
It may be very close to
the threshold of $N\Delta$. Similar to eq. (\ref{rwf1}), the relative
wave function between $N$ and $\Delta$ can also be expressed as
\eq \chi(r)_{N\Delta}=\sum_{m=1}^4\frac{c'_m}{\sqrt{4\pi}}\exp
\Big(-\frac{r^2}{2b_m^{'2}}\Big), \en where $c'_m$ and $b'_m$ can be
determined by fitting the relative wave function in this form to the
resultant wave function of the $D_{12}$ in our model calculation.\\

We first calculate the charge distribution of $D_{12}$ by assuming
it as a $N\Delta$ 6-quark system. The procedure is the same
as that for the $\Delta\Delta$ system shown in Sec. IIIA.
The obtained charge distribution can be written in the
following form:
\eq
G_E^{B}(Q^2;m_t)=3\Big [\hat{\cal
S}_{N}\hat{\cal E}^{m_t}_3+ \hat{\cal S}_{\Delta}\hat{\cal
E}^{m_t}_6\Big ],
\en
where $m_t$ stands for the third component of
the isospin of $D_{12}$ and
\eq
\hat{\cal S}_{N,~\Delta}(Q^2)=\exp\Big
[-\frac{b^2_{N,~\Delta}\vec{q}^{~2}}{6}\Big] \times {\cal
O}_{\chi_{N\Delta}}
\en
with
\eq
{\cal O}_{\chi_{N\Delta}}(Q^2)=<\chi_{N\Delta}(\vec{k}+\vec{q}/2)\mid\chi_{N\Delta}(\vec{k})>=
\sum_{mn}c'_mc'_n\frac{b_m^{'3}b_n^{'3}}{4\pi} \Big
(\frac{2\pi}{b_m^{'2}+b_n^{'2}} \Big )^{3/2}exp\big
[-\frac{b_m^{'2}b_n^{'2}}{8(b_m^{'2}+b^{'2}_n)}\vec{q}^2\big ],
\en
and
\eq
\hat{\cal E}^{m_t}_3&=& \langle~1,m_t
|~e_3~|~1,m_t~\rangle ~=~\frac16\Big
(\frac12,~1,~\frac32),~~~~~~~~~(m_t=1,0,-1)
\nonumber \\
\hat{\cal E}^{m_t}_6&=& \langle~1,m_t
|~e_6~|~1,m_t\rangle~=~\frac16\Big (\frac72,~1,~-\frac32\Big
),~~~~~(m_t=1,0,-1).
\en
\\ \noindent\par

Now, we calculate the charge distribution of $\ds$
which is assumed to have a $D_{12}\pi$ structure. What we would like
to see is that if such type of structure of $\ds$ has a
distinguishable charge distribution compared with the one shown
in Sec. IIIA. Since the $I(J)^{P}$ values of $D_{12}$ are $1(2)^+$,
the relative motion between $D_{12}$ and $\pi$ has to be at least P-wave,
and $D_{12}\pi$ in the isospin space should be decomposed
as
\eq\label{decom}
\mid \ds>=\frac{1}{\sqrt{3}}\Big
[D_{12}(I_z=1)\pi^--D_{12}(I_z=0)\pi^0+D_{12}(I_z=-1)\pi^+\Big ].
\en
The Jacobi momenta of $D_{12}$$-\pi$ system
are
\eq
\vec{{\cal
P}}=\vec{p}_{\pi}+\vec{p}_{_{D_{12}}},~~~~~\vec{\tilde
q}=\frac{m_{\pi}\vec{p}_{_{D_{12}}}-M_{_{D_{12}}}\vec{p}_{\pi}}{m_{\pi}+M_{_{D_{12}}}}
=a\vec{p}_{_{D_{12}}}-b\vec{p}_{\pi},
\en
where $\vec{p}_{\pi}$ and $\vec{p}_{_{D_{12}}}$
are the momenta of $\pi$ and $D_{12}$, and
$\vec{{\tilde q}}$ stands for the relative momentum between
the two systems. From the above equation, one sees
that the bombarding effect of photon on the relative
momentum $\vec{\tilde{q}}$ is much smaller in the case where
$D_{12}$ is stricken than that in the case where $\pi$ is hit. This
is because of the factors of $a=m_{\pi}/\Big
(M_{_{D_{12}}}+m_{\pi}\Big )\sim 6/100$ and
$b=1-a=M_{_{D_{12}}}/\Big (M_{_{D_{12}}}+m_{\pi}\Big )\sim 94/100$.\\

Although the relative wave function between $D_{12}$
and $\pi$ is not strictly solved, one can still qualitatively see a
general character of such a structure by assuming it being a
$P-$wave function
\eq
\chi_{_{D_{12}\pi}}(\tilde{q}^2,m_l)~=~\frac{\sqrt{2}b^{5/2}}{\pi^{3/4}}
{\mathcal{Y}}_{1m_l}(\Omega_{\tilde{q}} )
\exp(-\tilde{b}^2\vec{\tilde{q}}^{~2})/2
\en
where
$\mathcal{Y}_{1,m_l}(\Omega_{\tilde{q}})$ is a so-called solid
harmonics with $(l,m_l)=(1,m_l)$, and ${\tilde b}$ denotes the
size-parameter for the relative motion between the $D_{12}$ and
$\pi$. The size-parameter $\tilde{b}$ can be adopted in a large
range, for instance from $0.6\thicksim 6 fm$, and the real
relative $p-$wave function could be a combination of such a function
with different size-parameters. Then the contribution of the
relative wave function between $D_{12}$ and $\pi$ can be calculated
by
\eq
\mathcal{O}^{rel}_{D_{12},\pi}(\tilde{q},m_l)=
\langle~\chi_{_{D_{12}}\pi}^{\star}(\vec{k}+
\vec{\tilde{q}},m_l)~|~\chi_{_{D_{12}}\pi}(\vec{k},m_l)~\rangle,
\en
where the subscript $D_{12}$ and $\pi$ in
$\mathcal{O}^{rel}_{D_{12},\pi}$ denote the situations when a photon
hits $D_{12}$ and $\pi$, respectively.
The phenomenological monopole parametrization of the pion
charge distribution can be borrowed from
Ref.~\cite{Ame86}
\eq
\hat{S}_{\pi}=F_{\pi}(Q^2)=\frac{1}{1+({Q^2}/{\Lambda_{\pi}^2})},
~~~~\Lambda_{\pi}^2=0.462~GeV^2.
\en

Then, the charge distribution of $\ds$ in the
$D_{12}\pi$ scenario can be written as

\eq
G_{E,m_{\ds}}^{B}(Q^2) ~=~\frac{1}{3}\Big[
\sum_{m_t,m_l}~C_{s_{D_{12}}m_{D_{12}}~lm_l}^{J_{\ds}m_{\ds}}~~
\Big(\mathcal{G}_E^{D_{12}}(m_{t})\mathcal{O}^{rel}_{D_{12}}(aq,m_l)
+\hat{\mathcal{S}}_{\pi}\mathcal{O}^{rel}_{\pi}((1-a)q,m_l)\Big)\Big].
\en

It should be stressed that the pion contributions from the
first and the third terms of eq.~(\ref{decom}) canceled each other
and the one from the second term vanishes since it relates to
$\pi^0$. Therefore, the charge distribution of $d^*$ in this
scenario only comes from the contributions by $D_{12}$ and
the relative motion between
$D_{12}\pi$. Averaging over the initial states with various
magnetic quantum numbers of $\ds$, one finally obtain the
charged distribution of $d^*$
\eq \label{Gdp}
G_E^{B}(Q^2)=\frac12f_{D_{12}}(a^2\vec{q}^{~2})\times {\cal O}_{\chi_{N\Delta}} \Big
[\exp\big (-b_N^2\vec{q}^{~2}/6\big )+\exp\big (-b_{\Delta}^2\vec{q}^{~2}/6\big )\Big
],
\en
where
\eq
f_{D_{12}}(a^2\vec{q}^{~2})=\Big (1-\frac{a^2{\tilde
b}^2}{5}q^2\Big )\exp\big [-\frac{a^2{\tilde b}^2}{16}\vec{q}^{~2}\big ].
\en

\subsection{Numerical results in the two scenarios}

In our calculations, we take $b_N=b_{\Delta}=0.5~fm$, $b_c=0.45~fm$,
$b^r_c=0.45~fm$, and $\tilde b=1.2~fm$ as inputs. The probabilities
in eq. (13) are $\alpha^2\sim 0.31$ and $\beta^2\sim 0.69$. In Fig.
1 we plot the channel wave functions of the $\Delta\Delta$ in both
the single $\Delta\Delta$ channel (Scenario A1) and coupled
$\Delta\Delta+CC$ channel (Scenario A2) approximations,
respectively.  In terms of the same chiral SU(3) constituent quark
model, the wave functions of the $N\Delta$ system with $(I,S)=(1,2)$
can also been obtained by performing a bound state RGM calculation
with a set of slightly varied model parameters whose values are
still in a reasonable region. It is shown that in such a wave
function, the $^5S_2$ component dominates the $D_{12}$ state with a
fraction of about $94.47$\%, when the $N\Delta$ system is weakly
bound with a binding energy of $\epsilon=4.17~MeV$. The wave
function of $D_{12}$ is displayed in Fig. 1.  In terms of the wave functions
in the A1, A2, and $D_{12}$ cases, the root-mean-square radii ($rms$) of
the $d^*$ can be straightforward calculated. The obtained $rms$ are listed in
Table~\ref{tab:rms}.

\begin{table}[htbp]
\caption{The calculated $rms$ for $d^*(2380)$ in the scenarios A1
and A2 and for our $D_{12}$ (in units of $fm$).}
\begin{center}
\begin{tabular}{||c|c|c|c||}\hline
\multirow{1}{*}{Cases} &\multicolumn{2}{c|}{$d^*(2380)$} &\multirow{1}{*}{$D_{12}$}\\
\cline{2-3}   & A1 & A2 & \\ \hline
~~~~~~~~~~~~~~$rms~~~(fm)$~~~~~~~~~~~~~~
&~~~~~~~~~~~~~~1.09~~~~~~~~~~~~~~
 &~~~~~~~~~~~~~~0.78~~~~~~~~~~~~~~   &~~~~~~~~~~~~~~2.39~~~~~~~~~~~~~~\\
\hline
\end{tabular}
 \label{tab:rms}
\end{center}
\end{table}

\begin{figure}[htbp]
\begin{center}
\epsfig{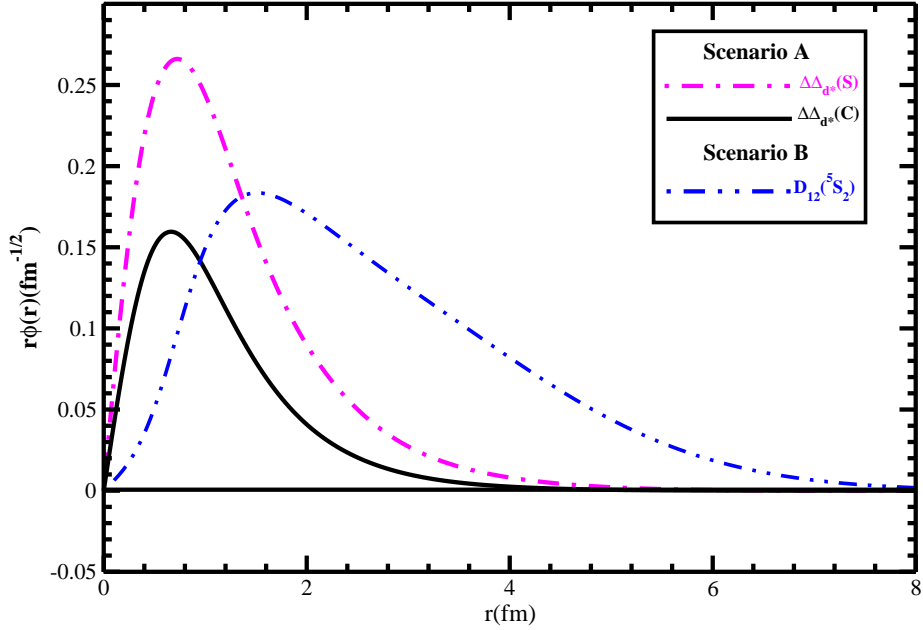}
\end{center}
\caption{\label{Fig1.} Wave functions. The pink dashed-dotted curve and the black solid curve
represent the $\Delta\Delta$ components in the $\ds$ wave function in the framework of
our chiral SU(3) constituent quark model with the single $\Delta\Delta$ (scenario A1)
channel only and the coupled $\Delta\Delta+CC$ channel (scenario A2), respectively.
The blue double-dotted-dashed curve shows the relative wave function
of the $^5S_2$ component in $D_{12}$ with binding energy of $\epsilon=4.17~\rm{MeV}$.}
\end{figure}

In Fig. 1 and Table~\ref{tab:rms}, one should notice that the
$\Delta\Delta$ wave function in the coupled channel approximation is
normalized to $0.31$. Clearly, we see that the wave function of
$D_{12}$ is much more extended in radius in comparison with that of
$\Delta\Delta$ does. This is reasonable, because the energy level of
this state is fairly close to the $N\Delta$ threshold, the system is
almost broken up, namely $N$ and $\Delta$ are "almost free" and the
separation between them becomes rather large. The smaller the
binding energy is, the larger the size of the system would be. The
root-mean-square radius of this component is $2.39~\rm{fm}$. This
value is consistent with the prediction from Heisenberg uncertainty
~\cite{Clement}.  \\

The charge distributions of the $d^*$ can be calculated by Eq.~(\ref{Gddcc})
in scenario A. The parameters $c_m$ and $b_m$ in Eq.~(\ref{rwf1}) can
be found in Refs.~\cite{Dong,Dong1}. The obtained charge distributions
from $\Delta\Delta$ and $CC$ components  are
demonstrated in Fig. 2. The charge distribution of the single
$\Delta\Delta$ model (scenario A1) is  plotted in Fig. 2 by a
pink dotted-dashed curve. The black solid and the red
dashed curves denote the contributions from the $\Delta\Delta$ and
$CC$ components (scenario A2), respectively, and the black dotted curve
represents the total contribution by summing over former two curves.
These curves tell us that the contribution from $CC$ component is larger
than that from the $\Delta\Delta$ component, especially, in the larger
momentum transfer region, the contribution is dominated by
the $CC$ component. It implies that the quark contents in the $CC$
component tend to concentrate in a more compact region than that in
the $\Delta\Delta$ component. This physics picture coincides with
the radii of two components calculated in our previous
paper~\cite{Huang}. Moreover, the curvature of
the distribution curve of A1 is larger than that in the coupled channel
case. It indicates that quarks here distribute in a larger region
than that in the coupled channel case. The size information of $\ds$
can also be seen from the slope of the distribution curve at the
origin, because such a slope is closely related to the radius of the system.
The larger the slope at origin is, the larger the radius of the system
would be. Comparing these slopes with the wave functions
shown in Fig. 1, we find that the obtained slopes at the origin here
coincide with the radii shown in Fig. 1. One sees that the radius
resulted from the single channel wave function (A1) is larger than one
from the two coupled-channel approximation (A2). Moreover, the radius
contributed by the $CC$ component is smaller than the one from $\Delta\Delta$
component.\\

In order to see the size character of $\ds$ with a $D_{12}-\pi$ structure,
the charge distribution of $d^*$ in the scenario B is calculated by Eq.~(\ref{Gdp}).
The obtained charge distribution curve, depicted by the blue double-dotted-dashed
curve, is also drawn in Fig. 2 as well. It should be specially mentioned that in our numerical
calculation, we do not solve the bound state problem for the $D_{12}\pi$ system,
explicitly. However, since the requirement of the conservations of the total spin
and parity, the relative motion between $D_{12}$ and $\pi$ must be at least a P-wave.
Therefore, if we ignore the component with higher partial wave, which will be greatly
suppressed, the true relative wave function would be a superposition of the P-wave wave
functions with different size-parameters $\tilde{b}$. Thus, we calculate the charge
distribution curves with a $\tilde{b}$ value from $0.6~fm$ to $6~fm$. The result shows
that those curves almost overlap with each other, namely this curve is almost insensitive
to the size-parameter $\tilde b$. This is because that the incoming photon is absorbed
by the $D_{12}$ system, and the induced change of the relative momentum between the
$D_{12}$ and $\pi$ is very small due to the factor of $m_{\pi}/\Big (M_{_{D_{12}}}+m_{\pi}\Big )\sim
6/100$.  Comparing the curve of scenario B with others of scenario A, we see that the curves
for the $D_{12}\pi$ scenario decrease and go to zero much faster than those for the scenario A.
A much larger slope at the origin means that the radius of the $D_{12}\pi$ system is much larger
in comparison with those in scenario A.  \\

From our numerical calculation, we find that the ratios of the
slopes of the curves at the origin in scenarios A1, A2, and B are
\eq 
R=\Big (-\frac{\partial G_E^{A1}(Q^2)}{\partial q^2}\Big )\Bigg
/ \Big (-\frac{\partial G_E^{A2}(Q^2)}{\partial q^2}\Big )\Bigg /
\Big (-\frac{\partial G_E^{B}(Q^2)}{\partial q^2}\Big )\Bigg / \Bigg
|_{Q^2=0}=2.30:1.45:6.86. 
\en 
It should be mentioned that the
contributions to $\frac{\partial G_E^{A1}(Q^2)}{\partial q^2}$ from
$\Delta\Delta$ and $CC$ components in the coupled channel
approximation of Scenario A2 are 0.64 and 0.81, respectively.   The
above obtained ratios tell us that the slop of the charge distribution 
of the $d^*$ in scenario B at $Q^2=0$  is about 4.8 times larger than 
the one in  Scenario A2, and the corresponding charge radius of scenario B
is about 2.2 times larger than the one in scenario A2. This feature is 
compatable with that discussed in Ref.~\cite{Gal1}. 
Finally, the very sharp charge distributions (or
large charge radii) of $d^*(2380)$ in the scenario B is mainly
dominated by the very broad wave function of the obtained $D_{12}$.
Therefore, we conclude that the two scenarios, A2 and B for the
$d^*(2380)$
give very different descriptions for its charge distribution and its charge radius.\\

\begin{figure}
\begin{center}
\epsfig{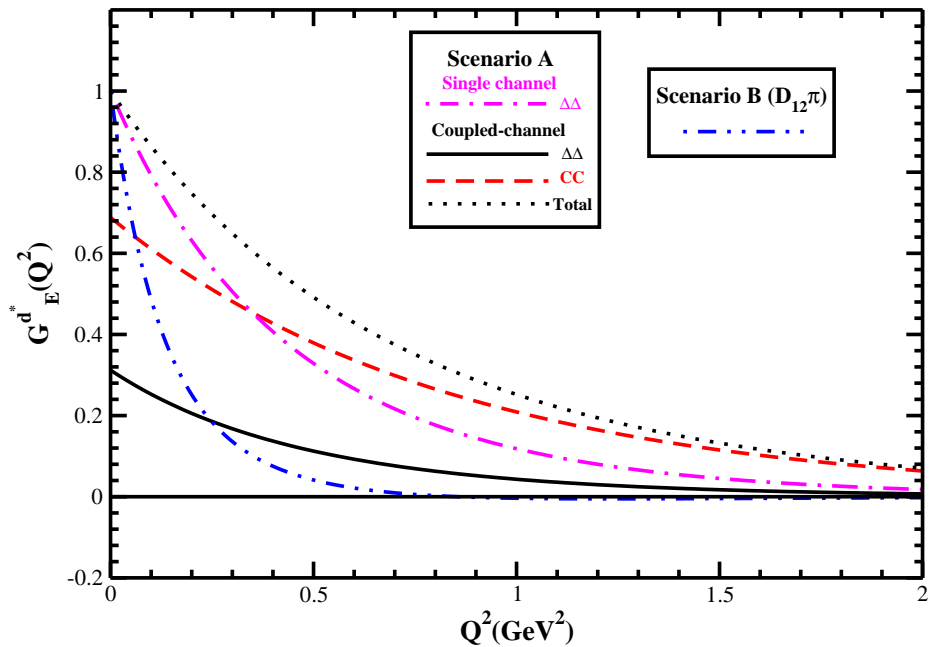}
\end{center}
\caption{\label{Fig2.}The obtained charge distributions
of $\ds$. The solid(black), red(dashed), and black(dotted) curves
stand for $\mid \Delta\Delta\chi_{\Delta}>$, $\mid cc\chi_c>$,
and the $total$ contributions in the coupled channel approximation,
respectively. The pink dotted-dashed curve represents the single
channel case of scenario A. The blue double-dotted-dashed curves
stand for the results of $D_{12}\pi$ scenario with $\epsilon\sim 4.17~\rm{MeV}$.}
\end{figure}

\section{Summary}

We have calculated the charge distribution of the $d^*$ in the two
scenarios; one is a hexaquark dominant picture and another is
a $D_{12}\pi$ resonant picture. In the first picture, we show the
total charge distribution and both contributions from its
$\Delta\Delta$ and $CC$ components. In order to make a comparison,
the result of the single $\Delta\Delta$ channel is also shown. Comparing the
predictions of the two scenarios, we see that the charge
distribution from the $D_{12}\pi$ system is remarkably different
from the scenario A2, and consequentially, the charge radius of the
the scenario B is obviously larger than that of the scenario A2.\\

We now expect a series of experiments which may
be able to test different interpretations of $d^*$
in future. Although the direct $ed^*$ scattering
measurement may be hard to carry out, one may consider the $d^*$
form factors in the time-like region. For example, the
production of the final $d^*\bar{d}^*$ pair in the $e^+e^-$ and
$p\bar{p}$ annihilation processes. It is our hope that the
future upgraded BEPC, Belle and Babar and experiment at $Panda$ with
high luminosity may provide a test for different theoretical
understandings.

\begin{acknowledgments}

We would like to thank Heinz Clement, Qiang Zhao, and Qi-Fang L\"u
for their useful and constructive discussions. This work is
supported by the National Natural Sciences Foundations of China
under the grant Nos. 11475192, 11475181, 11521505, 11565007, and
11635009, and by the fund provided to the Sino-German CRC 110
``Symmetries and the Emergence of Structure in QCD" project by
NSFC under the grant No.11621131001, the IHEP Innovation
Fund under the grant No. Y4545190Y2. F. Huang is grateful for the
support of the Youth Innovation Promotion Association of CAS under
the grant No. 2015358.
\end{acknowledgments}

\end{document}